%% file: main.tex
\documentclass[10pt,twocolumn,letterpaper]{article}

\usepackage[pagenumbers]{wacv} 
\usepackage{multirow}%
\usepackage{graphicx}
\usepackage{amsmath}
\usepackage{booktabs}
\usepackage{float}
\usepackage[utf8]{inputenc} 
\usepackage[T1]{fontenc}    
\usepackage{url}            
\usepackage{booktabs}       
\usepackage{amsfonts}       
\usepackage{nicefrac}       
\usepackage{microtype}      
\usepackage{xcolor}         
\usepackage{float}
\usepackage[linesnumbered,ruled,vlined]{algorithm2e}

\RestyleAlgo{ruled} 
\LinesNumbered
\usepackage{listings}%
\usepackage{pgfplots}%
\usepackage{tikz}
\usetikzlibrary{shapes.geometric, arrows}
\usepackage{url}


\definecolor{wacvblue}{rgb}{0.21,0.49,0.74}
\usepackage[pagebackref,breaklinks,colorlinks,allcolors=wacvblue]{hyperref}
\def\wacvPaperID{2349} 
\def\confName{WACV}
\def\confYear{2026}

\title{From Attention to Frequency: Integration of Vision Transformer and FFT-ReLU for Enhanced Image Deblurring}

\author{
Syed Mumtahin Mahmud, 
Mahdi Mohd Hossain Noki, 
Prothito Shovon Majumder, \\
Abdul Mohaimen Al Radi, 
Md. Haider Ali,
Md. Mosaddek Khan \\\\
Department of Computer Science and Engineering, University of Dhaka, \\
Dhaka-1000, Bangladesh\\
{\tt\small \{syedmumtahin-2019617833, mahdimohdhossain-2019017785, prothitoshovon-2018725302,}\\
{\tt\small abdulmohaimenal-2018925300\}@cs.du.ac.bd}, \tt\small\{haider, mosaddek\}@du.ac.bd
}

\begin{document}
\maketitle
\input{sec/0_abstract}    
\input{sec/1_intro}
\input{sec/2_relatedwork}

\input{sec/3_methodology}

\input{sec/4_experiment}
\input{sec/5_conclusion}

{
    \small
    \bibliographystyle{ieeenat_fullname}
    \bibliography{main}
}

\end{document}

%% file: sec/0_abstract.tex
\begin{abstract}
Image deblurring is vital in computer vision, aiming to recover sharp images from blurry ones caused by motion or camera shake. While deep learning approaches such as CNNs and Vision Transformers (ViTs) have advanced this field, they often struggle with complex or high-resolution blur and computational demands. We propose a new dual-domain architecture that unifies Vision Transformers with a frequency-domain FFT-ReLU module, explicitly bridging spatial attention modeling and frequency sparsity. In this structure, the ViT backbone captures local and global dependencies, while the FFT-ReLU component enforces frequency-domain sparsity to suppress blur-related artifacts and preserve fine details. Extensive experiments on benchmark datasets demonstrate that this architecture achieves superior PSNR, SSIM, and perceptual quality compared to state-of-the-art models. Both quantitative metrics, qualitative comparisons, and human preference evaluations confirm its effectiveness, establishing a practical and generalizable paradigm for real-world image restoration.\footnote{The source code is available in the authors' GitHub profile.}
\end{abstract}

%% file: sec/1_intro.tex
\section{Introduction}
\label{sec1}
Image deblurring is a crucial task in image processing, aiming to recover sharp images from blurry ones caused by factors like camera shake or motion blur. Computer vision research has recently seen significant attention focused on the fundamental challenges of image restoration. As the number of individuals capturing photographs using their smartphones and other portable cameras continues to rise, this issue is gaining greater importance.  The process of reconstructing a sharp image from its blurry input is very complex and challenging. Dealing with extreme blur or complicated images is one area where traditional approaches often struggle to provide adequate results.

Traditional image deblurring methods \citep{articleWang,articleDonatelli, Xu, articlekernel} rely on explicit blur kernel estimation and strong priors. While mathematically elegant, these approaches assume uniform blur and are highly sensitive to noise, making them unsuitable for complex real-world conditions. Assuming that the blur is uniform across the entire image is an approach to simplifying the issue. But these techniques are very noise-sensitive, and the quality of the image might deteriorate, and major artifacts (such as ringing or checkerboard patterns, especially near edges or sharp transitions in the image) can arise from even a tiny quantity of noise.

Recent deep learning advances introduced CNN-based methods for deblurring \citep{9044873cnn1,8099518cnn2,gao2019dynamiccnn2,9578045}. They effectively handle complex blur patterns, are robust to noise, and benefit from large datasets through end-to-end learning with multi-scale features. However, CNNs struggle with long-range dependencies, require extensive labeled data, and often fail to generalize to unseen distortions. They are also sensitive to changes in scale, aspect ratio, or image quality, limiting adaptability. Following the success of Transformers in NLP tasks \citep{vaswani2017attentionisalluneed}, Vision Transformers (ViTs) \citep{vsionTransformer} have emerged as strong alternatives to CNNs by capturing long-range dependencies and modeling global context. They provide scalability, flexibility, and reduced inductive bias, making them effective for complex vision problems. However, applying ViTs to high-resolution deblurring is computationally expensive, and they may struggle with generalization to unseen distortions, over-reliance on data, and issues such as overfitting, artifacts, and limited interpretability \citep{Han2020ASO,lim}.

Attempts to combine blind deconvolution with deep learning \citep{mao2023intriguing,10040651Yae} remain limited, as they fail to capture the joint spatial–frequency complexity of real blur. Existing approaches either operate purely in the spatial domain (CNNs, Transformers) or rely solely on frequency manipulations (FFT-based deblurring). In contrast, our work is the first to present a tightly coupled hybrid model that combines both. This allows us to capture global spatial dependencies while also correcting blur at the frequency level, bridging two historically separate paradigms in image restoration. In the wake of the above background, the summary of the specific contributions made by this paper is listed below:

\begin{enumerate}[1.]

\item We propose a principled dual-domain paradigm that explicitly integrates the spatial domain (Vision Transformer) with the frequency domain (FFT-ReLU). Unlike prior approaches that remained domain isolated, operating either in the spatial domain or solely in the frequency domain, our method unifies both in a complementary design. By bridging two historically separate restoration strategies, the model achieves robust deblurring across structured motion blur while also recovering fine high-frequency details.

\item We compare our proposed method against state-of-the-art image deblurring models on benchmark datasets. Our proposed approach provides sharper, more compelling, and visually convincing images. 

\item We conduct an experiment, Human Visual Preference Evaluation, along with computational metrics, helps us assess the effectiveness of image deblurring. This human-centered approach complements quantitative measures by offering insight into the practical performance of our technique from a Human-Computer Interaction standpoint.
\end{enumerate}

%% file: sec/2_relatedwork.tex
\section{Related Works}  \label{related_works}
Image restoration and enhancement have witnessed substantial advancements in recent years, primarily driven by the application of deep learning methodologies. Convolutional Neural Networks (CNNs) established the foundation for many initial approaches \citep{Dong2016, Zhang2018}, demonstrating the efficacy of learned features for these tasks. Subsequently, more sophisticated architectures emerged, such as multi-scale encoder-decoder networks \citep{tao2018srndeblur}, which facilitated the integration of both local and global image features.
The advent of attention mechanisms and Transformer architectures \citep{liang2021swinir, Restormer} represented a significant progression, enabling models to capture long-range dependencies in images with greater effectiveness. However, the increased computational requirements of these models have prompted a parallel line of research focused on computational efficiency \citep{Chen2022}.

In the domain of image deblurring, numerous contemporary approaches concentrate predominantly on spatial domain processing. However, recent investigations have begun to explore the potential of frequency domain techniques. For instance, \citep{Yang2020} proposed Fourier domain adaptation for semantic segmentation, while \citep{Rao2021} developed global filter networks for image classification. These studies demonstrate the potential utility of frequency domain information in various image processing tasks.

Building upon these advancements, recent research has explored hybrid models that combine the strengths of CNNs and Transformers \citep{app13010311}. However, these hybrid models also come with their own set of challenges. One notable drawback is the increased computational complexity and resource requirements. Combining CNNs and Transformers can lead to higher memory usage and longer training times due to the need to process both local and global features simultaneously. Recent work on FFT-ReLU sparsity priors has demonstrated that frequency-domain constraints can enable efficient and competitive blind deblurring across diverse image types \cite{radi}. However, such approaches lacked the ability to capture complex spatial dependencies, limiting their robustness for spatially varying blur. In this paper, we extend the scope of FFT-ReLU by integrating it with Vision Transformers, creating an explicit spatial–frequency framework. This hybrid design combines the efficiency of frequency-domain priors with the representational strength of deep learning, resulting in a more generalizable and practical solution for real-world deblurring.


%% file: sec/3_methodology.tex
\begin{figure*}[t]
\centering
\includegraphics[width=\linewidth]{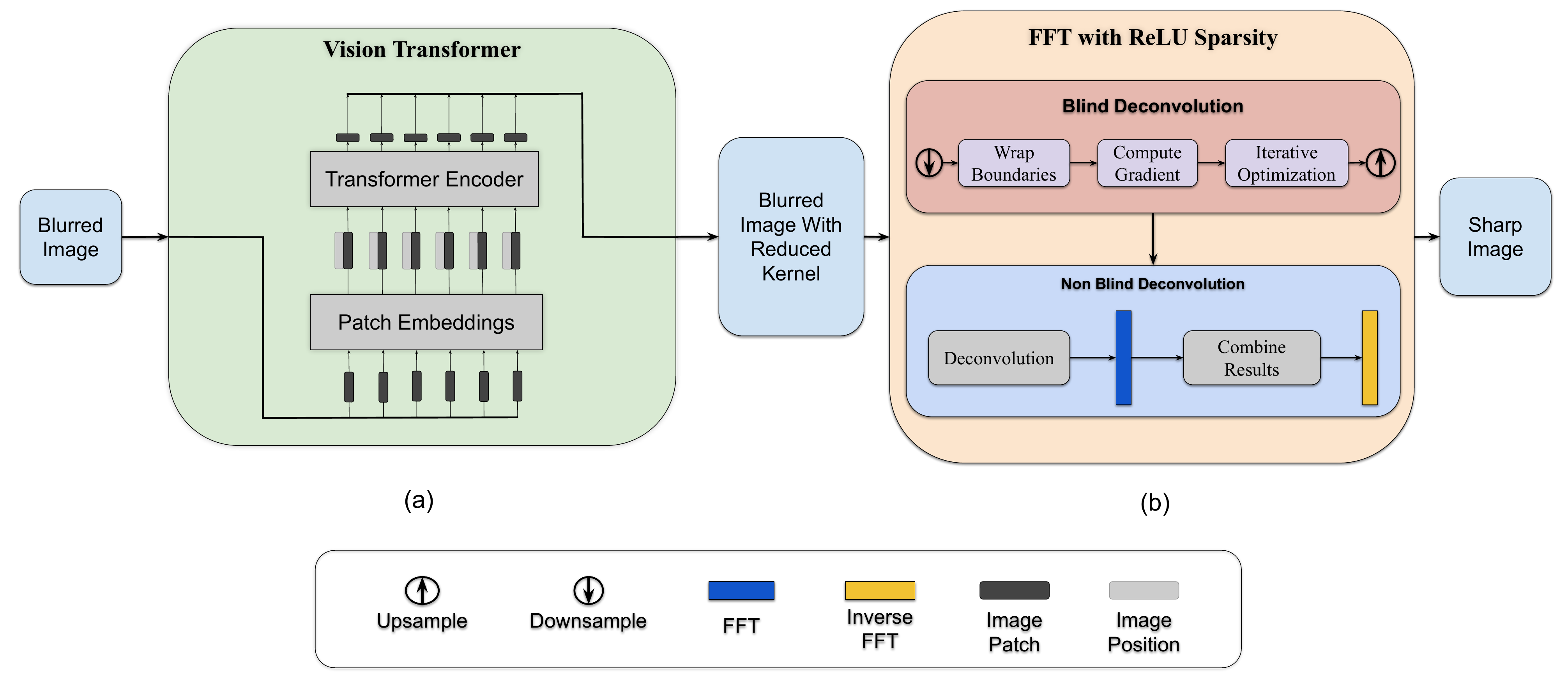}
\caption{ (a) Uses a Vision Transformer to extract features from blurred images and reduce the blur kernel for further processing. (b) applies FFT-based blind and non-blind deconvolution, utilizing ReLU sparsity, to restore a sharp image}
\label{fig:diagram}
\end{figure*}
\section{Methodology}\label{Methodology}
We perform deblurring in two phases. First, a Vision Transformer model preprocesses the image and narrows down the possible blur kernel values. In the second phase, FFT-based blind and non-blind deconvolution is applied. Unlike existing approaches that operate exclusively in the spatial domain (CNNs, Transformers) or solely in the frequency domain (FFT-based methods), our methodology unifies both, allowing the spatial module to reduce kernel ambiguity while the frequency module suppresses artifacts and restores fine detail. This explicit dual-domain integration is what makes our design robust and practically effective. Together, they form a complementary system where each domain addresses the other’s blind spots. Importantly, the FFT-ReLU stage is highly parallelizable on GPUs, making the dual-domain design practical for high-resolution images without prohibitive overhead.

\subsection{Vision Transformer as a Pre-processing Module}
In our approach, we leverage the Transformer's self-attention mechanism to model relationships between elements in the input image, which allows us to capture both local and global dependencies effectively. We begin by embedding the input image into a high-dimensional space, enhanced with positional encodings to retain spatial order. Through the Transformer's multi-head self-attention, we are able to weigh the importance of various parts of the image. These attention outputs are then processed through feed-forward layers with residual connections, allowing us to extract deep hierarchical features. This process is crucial in limiting the domain of the blur kernel.

\subsection{FFT With ReLU Sparsity}

In this section, we explain the blind and non-blind deconvolution techniques used in our image deblurring approach. Blind deconvolution aims to recover a sharp image when the blur kernel is unknown, making it a challenging, ill-posed problem. Once the blur kernel is estimated, non-blind deconvolution refines the image by using the known kernel to enhance sharpness. Our method combines Total Variation denoising and L0 gradient minimization to achieve high-quality results, while optimized padding reduces boundary artifacts.

\subsubsection{Blind Deconvolution}

Blind deconvolution is a cornerstone of our image restoration framework, designed to recover a sharp image from a blurred one without prior knowledge of the blur kernel, also known as the Point Spread Function (PSF). This task is inherently challenging as it requires the simultaneous estimation of both the original sharp image and the blur kernel, rendering it an ill-posed problem in the realm of image processing. We address this challenge through an iterative optimization-based blind deconvolution algorithm, detailed in Algorithm~\ref{alg:Blind Deconvolution}. Our approach uses $L_0$-gradient minimization, which works effectively to estimate both the latent sharp image and the blur kernel. Below is an in-depth description of our blind deconvolution algorithm.

\begin{algorithm}[t]
\caption{Blind Deconvolution}
\label{alg:Blind Deconvolution}
\KwIn{Blurred image $B$, initial kernel estimate $k$, parameters $\lambda_{ftr}$, $\lambda_{grad}$, threshold, opts}
\KwOut{Deblurred image $S$, estimated kernel $k$}

$B_w \gets \text{wrap\_boundary\_liu}(B)$\;\label{a1l1}
$B_x, B_y \gets \text{compute\_gradients}(B_w)$\;\label{a1l2}

\For{$iter = 1$ \KwTo $opts.xk\_iter$}{
    $S \gets \text{$L_0Deblur\_FTR$}(B, k, \lambda_{ftr}, \lambda_{grad})$\;
    $latent_x, latent_y, threshold \gets \text{threshold\_pxpy\_v1}(S, k.size, threshold)$\;
    $k_{prev} \gets k$ \label{a1l6}\;
    $k \gets \text{estimate\_psf}(B_x, B_y,$ $latent_x, latent_y, weight, k_{prev}.size)$\label{a1l7}\;
    $k \gets \text{prune\_isolated\_noise}\;(k)$\; \label{a1l8}
    $k \gets \text{normalize}(k)$\;\label{a1l9}
    Update $\lambda_{grad}, \lambda_{ftr}$ \label{a1l10}
}
\Return{$k, S$}
\end{algorithm}

Our algorithm initiates the process by preparing the blurred image using the \texttt{wrap\_boundary\_liu} function (Line~\ref{a1l1}), which effectively handles boundary conditions to prevent artifacts in subsequent processing steps. Following this, we compute the gradients of the wrapped image ($B_x$ and $B_y$)(Line~\ref{a1l2}), which are essential for the iterative optimization process. We employ an alternating strategy to estimate the latent sharp image and the PSF. For latent image estimation, we invoke the $L_0$Deblur\_FTR method.

Once the latent image $S$ is estimated, the algorithm proceeds to PSF estimation. (Line~\ref{a1l7}), which leverages the computed gradients and the current estimate of the latent image to update the blur kernel $k$. Following the estimation, we refine the kernel by removing isolated noise via the \texttt{prune\_isolated\_noise} function (Line~\ref{a1l8}) and subsequently normalize it to ensure consistency (Line~\ref{a1l9}). We use these refinement steps to achieve an accurate and clean PSF estimation. A standout feature of our algorithm is the adaptive adjustment of regularization parameters ($\lambda_{\text{grad}}$, $\lambda_{\text{ftr}}$) during each iteration (Line~\ref{a1l10}). This progressive tuning facilitates the restoration of increasingly finer image details as the deconvolution process advances, culminating in a sharper and more precise final image.

In the following sections, we describe the core procedures integral to our blind deconvolution algorithm.

\subsubsection*{PSF Estimation} PSF estimation, a process that identifies the point spread function responsible for image blur, is a pivotal component of our blind deconvolution process. We utilize Fourier domain operations to enhance computational efficiency, which is particularly advantageous for processing large-scale images. The procedure commences by transforming the PSF from the spatial domain to the Optical Transfer Function (OTF) in the frequency domain. This transformation accelerates subsequent computations by exploiting the properties of frequency domain operations. After performing the necessary multiplications and manipulations in the frequency domain, we convert the result back to the spatial domain for further processing. To stabilize the solution and mitigate overfitting, we incorporate a Tikhonov regularization term ($\lambda * x$) into our cost function. This regularization plays a crucial role in addressing the ill-posed nature of the blind deconvolution problem, ensuring a robust and reliable PSF estimation.

\subsubsection*{$L_0$Deblur-FTR}
\begin{algorithm}[t]
\setcounter{algocf}{0}
\SetAlgorithmName{Procedure}{}{}
\caption{$L_0$Deblur\_FTR}
\label{proc:L0Deblur_FTR}
\KwIn{Image $Im$, kernel $k$, $\lambda$, $wei\_grad$, $\kappa$}
\KwOut{Deblurred image $S$}

$S \gets \text{wrap\_boundary\_liu}(Im)$ \label{p1l1}

Initialize FFT matrices and parameters\;

\While{$\alpha < \alpha_{max}$}{\label{p1l3}
    $ftr \gets \text{findM}(S)$ \label{p1l4}
    
    $Mat \gets \text{threshold}(ftr, \mu, \alpha)$\;

    \While{$\beta < \beta_{max}$}{  \label{p1l6}
        Update $S$ using FFT operations and gradients\;
        $\beta \gets \beta * \kappa$ \label{p1l8}
    } \label{p1l9}
    $\alpha \gets \alpha * \kappa$ \label{p1l10}
}\label{p1l11}
$S \gets \text{crop\_to\_original\_size}(S)$\;
\Return{$S$}
\end{algorithm}

The $L_0$Deblur\_FTR method incorporates the Fast Total Variation Regularization (FTR) for latent image estimation, as detailed in Procedure~\ref{proc:L0Deblur_FTR}. FTR is renowned for its ability to preserve edges while effectively smoothing out noise in images.

The algorithm commences with an efficient boundary handling step using the \texttt{wrap\_boundary\_liu} function (Line~\ref{p1l1}), which is essential for preventing artifacts at the image edges during FFT operations. The $L_0$Deblur\_FTR method employs an iterative optimization scheme with a nested loop structure. The outer loop continues until the threshold parameter $\alpha$ reaches its maximum value (Lines~\ref{p1l3} to ~\ref{p1l11}), while the inner loop updates the sharp image estimate (Lines~\ref{p1l6} to \ref{p1l9}).

A key component of this method is the \texttt{findM} procedure (Line~\ref{p1l4}), which computes specific matrices used in the regularization process, enhancing the method's ability to preserve edges while mitigating noise. The gradual increase of parameters $\alpha$ and $\beta$ implements a coarse-to-fine strategy, allowing the algorithm to capture both large-scale structures and fine details within the image. Upon completion of the iterative process, we crop the deblurred image back to its original size (Line~\ref{p1l11}) to finalize the restoration. The algorithm utilizes some utility methods, like the \texttt{psf2otf} converter, to transform the PSF to its corresponding OTF in the Fourier domain, significantly accelerating the processing of large images. Additionally, it adopts a coarse-to-fine optimization strategy, where regularization parameters are incrementally increased to enhance the quality of the deblurred image progressively. Furthermore, it implements alternating minimization, iterating between the estimation of the latent sharp image and the gradient field, ensuring that both the image and its gradients are refined in tandem for optimal deblurring results.

In summary, our blind deconvolution algorithm alternates between estimating the latent sharp image and the PSF, employing adaptive regularization techniques to achieve effective deblurring results. 

\begin{algorithm}[h]
\setcounter{algocf}{1}
\caption{Ringing Artifact Removal}
\label{alg:ringing_artifact_removal}
\KwIn{Blurry image $y$, blur kernel $k$, $\lambda_{tv}$, $\lambda_{L_0}$, $weight_{ring}$}
\KwOut{Deblurred image with reduced ringing artifacts}

$H, W, C \gets \text{dimensions of } y$ \; \label{a4l1}
$p \gets \texttt{opt\_fft\_size}([H + k.\text{shape}[0] - 1, W + k.\text{shape}[1] - 1])$ \; \label{a4l2}
$y_{pad} \gets \texttt{wrap\_boundary\_liu}(y, p)$ \; \label{a4l3}
$Latent_{tv} \gets \text{empty tensor}$\;

\For{$c = 1$ \KwTo $C$}{ \label{a4l5}
    $aniso \gets \texttt{deblurring\_adm\_aniso}(y_{pad}[:,:,c], k, \lambda_{tv}, 1)$\;
    $Latent_{tv} \gets \text{concatenate}(Latent_{tv}, aniso)$ \; \label{a4l7}
} \label{a4l8}
$Latent_{tv} \gets Latent_{tv}[:H, :W, :]$ \; \label{a4l9}

\If{$weight_{ring} = 0$}{ \label{a4l10}
    \Return $Latent_{tv}$\;
}

$Latent_{L_0} \gets \texttt{L$_0$Restoration}(y_{pad}, k, \lambda_{L_0}, 2)$ \; \label{a4l12}
$Latent_{L_0} \gets Latent_{L_0}[:H, :W, :]$ \; \label{a4l13}
$diff \gets Latent_{tv} - Latent_{L_0}$ \; \label{a4l14}
$bf\_diff \gets \texttt{bilateral\_filter}(diff, 3, 0.1)$ \; \label{a4l15}
$result \gets Latent_{tv} - weight_{ring} \cdot bf\_diff$ \; \label{a4l16}
\Return $result$ \; \label{a4l17}
\end{algorithm}

\subsubsection{Non-Blind Deconvolution} The blind deconvolution is followed by a non-blind deconvolution to find a sharp image from a blurred image, given the blur kernel produced in the previous procedure. We leverage a combination of Total Variation (TV) denoising and L$_0$ gradient minimization to produce a high-quality deblurred image in the Algorithm \ref{alg:ringing_artifact_removal}. It takes as input a blurry image $y$, an estimated blur kernel $k$, and parameters $\lambda_{\text{tv}}$, $\lambda_{\text{L$_0$}}$, and $weight_{ring}$ that control the deblurring process. The dimensions of the input image $y$ are obtained (Line \ref{a4l1}), and the padding size $p$ is calculated using an optimized FFT size function (Line \ref{a4l2}). This padding is applied to reduce boundary artifacts in Fourier-based deconvolution, using a method specifically designed for this purpose (Line \ref{a4l3}).

\subsubsection*{Total Variation (TV) Denoising} Next, we apply TV denoising to each color channel separately (Lines \ref{a4l5} to \ref{a4l8}). TV denoising promotes piecewise smooth solutions, effectively preserving edges while removing noise and small-scale artifacts. We achieve this by iteratively updating the image estimate and promoting sparsity in the gradient domain through the Alternating Direction Method (ADM). The denoised channels are concatenated to form the latent image $Latent_{tv}$ (Line \ref{a4l7}). The final TV-denoised result is then extracted to match the original image dimensions (Line \ref{a4l9}).

\subsubsection*{Ringing Artifact Handling} If the parameter $weight_{ring}$ is zero, the algorithm directly returns the TV-denoised result (Line \ref{a4l10}). Otherwise, the algorithm proceeds to apply L$_0$ gradient minimization using the \texttt{L$_0$Restoration} function (Line \ref{a4l12}). L$_0$ minimization aims to produce a sharp image by minimizing the number of non-zero gradients, which is particularly effective for removing small-scale artifacts and enhancing edge sharpness. The L$_0$-minimized result is then cropped to match the original image dimensions (Line \ref{a4l13}).

\subsubsection*{Final Image Combination} To leverage the strengths of both TV and L$_0$ methods, the algorithm computes the difference between the TV and L$_0$ results (Line \ref{a4l14}). This difference is processed with a bilateral filter, which is an edge-preserving smoothing filter (Line \ref{a4l15}). The filtered difference is weighted by $weight_{ring}$ and subtracted from the TV result to produce the final deblurred image (Line \ref{a4l16}). This result, with reduced ringing artifacts, is returned as the output of the algorithm (Line \ref{a4l17}).

In summary, our dual-domain design enhances deblurring by combining spatial attention with frequency sparsity, enabling sharper detail recovery and reduced artifacts. This integration improves robustness across diverse blur patterns and delivers perceptual gains that align with real-world needs.

%% file: sec/4_experiment.tex
\section{Experiments and Results}\label{Results}
In this section, we examine the effectiveness of our suggested model by analyzing the experimental findings. Initially, we describe the benchmark datasets that we have used for the experiment. Next, we compare it quantitatively and qualitatively with the current state-of-the-art approaches.

\subsection{Pre-Processor}
Restormer\footnote{Any Vision Transformer can be tailored to be used as a pre-processing step in place of Restormer} was selected as the Transformer model for our image deblurring methodology due to its demonstrated efficacy in capturing long-range dependencies within images while maintaining computational efficiency. Its architecture is particularly well-suited for high-resolution image processing, which is critical for tasks such as image deblurring that require the precise capture of both local and global image features. A key advantage of Restormer lies in its multi-head self-attention mechanism, which facilitates the modelling of complex, long-range interactions across various spatial regions of an image \citep{Restormer}.

\subsection{Datasets Details and Experimental Setup}
For producing results and comparisons, we have used the GoPro \citep{Gopro}, HIDE \citep{Hide}, RealBlur \citep{Realblur} and Kohler \citep{kohler} datasets. These datasets are widely used for image deblurring training and testing purposes. The results are then compared quantitatively and qualitatively with other state-of-the-art deblurring models. Our experiments and model training were conducted on a computer configuration that was designed for optimal performance. It included an NVIDIA GeForce RTX 4090 GPU with 24GB of GDDR6X memory, which was accompanied by an AMD Ryzen 9 5950X CPU with 16 cores (32 threads) functioning at 3.4GHz. This hardware configuration enabled  Restormer to be trained effectively on the GoPro dataset.

\subsection{Result Comparison}

In this section, we evaluate the performance of our proposed method against several state-of-the-art approaches to demonstrate both its quantitative and qualitative strengths. We begin by presenting a detailed analysis of numerical metrics before moving on to perceptual and user-based evaluations

\subsubsection{Quantitative Results}
\label{Quantitative Results}
\begin{table*}
\setlength\tabcolsep{0pt}
    \centering
    \footnotesize
    \caption{Quantitative evaluations of the proposed method against state-of-the-art methods on the benchmark RealBlur-J, RealBlur-R, HIDE, and GoPro datasets. Those with the highest PSNR/SSIM scores in each column are \textbf{highlighted}}\label{result-compasiron}%
    \begin{tabular*}{\textwidth}{@{\extracolsep{\fill}}lcccccccccccc}
        \toprule
        \multirow{2}{*}{\textbf{Model Name}} & \multicolumn{2}{c}{\textbf{RealBlur-J}} & \multicolumn{2}{c}{\textbf{RealBlur-R}} & \multicolumn{2}{c}{\textbf{HIDE}} & \multicolumn{2}{c}{\textbf{GoPro}} & \multicolumn{2}{c}{\textbf{Average}} \\
        \cmidrule(r){2-3} \cmidrule(r){4-5} \cmidrule(r){6-7} \cmidrule(r){8-9} \cmidrule(r){10-11}
        & \textbf{PSNR} & \textbf{SSIM} & \textbf{PSNR} & \textbf{SSIM} & \textbf{PSNR} & \textbf{SSIM} & \textbf{PSNR} & \textbf{SSIM} & \textbf{PSNR} & \textbf{SSIM} \\
        \midrule
        Restormer & 28.96 & 0.879 & 36.19 & \textbf{0.957} & \textbf{31.22} & \textbf{0.942} & 32.92 & 0.961 & 32.3225 & 0.93475 \\
        MPRNet & 28.7 & 0.873 & 35.99 & 0.952 & 30.96 & 0.939 & 32.66 & 0.959 & 32.0775 & 0.93075 \\
        Stripformer & 28.82 & 0.876 & 36.08 & 0.954 & 31.03 & 0.939 & \textbf{33.08} & \textbf{0.962} & 32.2525 & 0.93275 \\
        NAFNet & \textbf{29.34} & \textbf{0.882} & 35.01 & 0.928 & 19.38 & 0.62 & 26.76 & 0.898 & 27.6225 & 0.832 \\
        SRN & 28.56 & 0.867 & 35.66 & 0.947 & 28.36 & 0.915 &  30.26 & 0.934 & 30.71 & 0.916 \\
        SPAIR & 28.81 & 0.875 & - & - & 30.29 & 0.931 & 32.06 & 0.953 & - & - \\ 
        MT-RNN & 28.44 & 0.862 & 35.79 & 0.951 & 29.15 & 0.918 & 31.15 & 0.945 & 31.13 & 0.919 \\ 
        DeblurGAN & 27.97 & 0.834 & 33.79 & 0.903 & 24.51 & 0.871 & 28.7 & 0.858 & 28.7425 & 0.8665 \\
        DeblurGANv2 & 28.7 & 0.866 & 35.26 & 0.944 & 26.61 & 0.875 & 29.55 & 0.934 & 30.03 & 0.90475 \\
        DBGAN & 24.93 & 0.745 & 33.78 & 0.909 & 28.94 & 0.915 & 31.1 & 0.942 & 29.6875 & 0.87775 \\
        DMPHN & 28.42 & 0.86 & 35.70 & 0.948 & 29.09 & 0.924 & 31.2 & 0.94 & 29.57 & 0.918 \\
        MIMO-UNet+ & 27.63 & 0.837 & 35.54 & 0.947 & 29.99 & 0.930 & 32.45 & 0.957 & 31.40 & 0.917 \\
        DeblurDiNAT-S &  28.80 & 0.877 & 35.92 & 0.954 & 30.65 & 0.936 & 32.85 & 0.961 & 32.05 & 0.932 \\
        
        \midrule
        Ours & 28.99 & 0.88 & \textbf{36.21} & \textbf{0.957} & \textbf{31.22} & \textbf{0.942} & 32.92 & 0.961 & \textbf{32.335} & \textbf{0.935} \\
        \bottomrule
    \end{tabular*}
    \label{tab:performance}
\end{table*}

We compared our proposed method with several state-of-the-art methodologies including Restormer \citep{Restormer}, MPRNet \citep{Zamir2021MPRNet}, Stripformer \citep{Tsai2022Stripformer}, NAFNet \citep{chu2022nafssrNafnet}, DeblurGAN \citep{DeblurGAN}, DeblurGANv2 \citep{DeblurGANv2}, DBGAN \citep{DBGAN}, DMPHN \citep{zhang2022eventDMPHN}, MIMO-UNet+ \citep{MIMO-UNet+}, DeblurDiNAT-S \citep{liu2024deblurdinat}, SRN \citep{tao2018srndeblur}, SPAIR \citep{spair}, MT-RNN \citep{park2020mtrnn}. 

According to Table \ref{result-compasiron}, our proposed methodology generates superior results in terms of PSNR and SSIM metrics when contrasting the models on the RealBlur-R and HIDE datasets. The RealBlur-J and GoPro datasets yield results that are nearly identical to the best available result. By taking the average performance across the GoPro, HIDE, RealBlur-R, and RealBlur-J datasets, our model demonstrates the best results, further solidifying its robustness and effectiveness in diverse scenarios. Our qualitative comparison results may indicate a slight improvement, mostly due to measurements like PSNR and SSIM being evaluated against a ground truth image. Although our technique may sometimes provide sharper images, this does not consistently correspond with the mathematical models used by PSNR and SSIM. Improved image sharpness may improve visual quality for human observers. However, it might result in a marginal reduction in metric values owing to the presence of heightened high-frequency features that do not align precisely with the ground truth. This discrepancy highlights a known limitation of pixel-based metrics and underscores the importance of perceptual evaluation. We have presented a quantitative comparison of the models on images from various datasets in Section \ref{Qualitative Results}. Given the impressive nature of our findings, we opted to conduct a human visual preference evaluation test (in Section \ref{HCI}), and it confirms that the improvements correspond to genuine perceptual gains.

\subsubsection{Qualitative Results}
\label{Qualitative Results}

We compared our proposed method qualitatively with DMPHN \citep{zhang2022eventDMPHN}, MPRNet \citep{Zamir2021MPRNet}, Stripformer \citep{Tsai2022Stripformer}, NAFNet \citep{chu2022nafssrNafnet} and along with Restormer \citep{Restormer}. We have added the images produced by each model along with the input (blurry image) and ground truth from each of the benchmark datasets (GoPro, HIDE, RealBlur-R and RealBlur-J). As most of the image deep learning image deblurring methods are trained and tested with these benchmark datasets, we sought to enhance the robustness and generalizability of our comparison by incorporating the Kohler \citep{kohler} dataset (in Figure \ref{fig:kohler1}. The Kohler dataset is well-regarded in the field of image deblurring for its challenging and realistic blur patterns.

\begin{figure}[h]
\centering
\includegraphics[width=\linewidth]{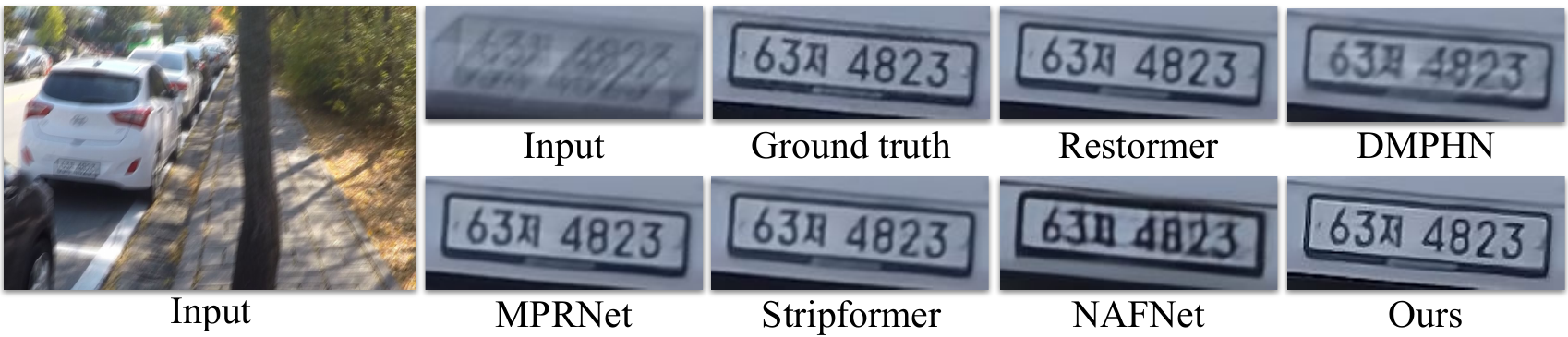}
\caption{Qualitative image deblurring comparison between state-of-the-art models and our method on GoPro \citep{Gopro}. Best viewed on a high-definition monitor when zoomed in}
\label{fig:Gopro}
\end{figure}
\begin{figure}[h]
\centering
\includegraphics[width=\linewidth]{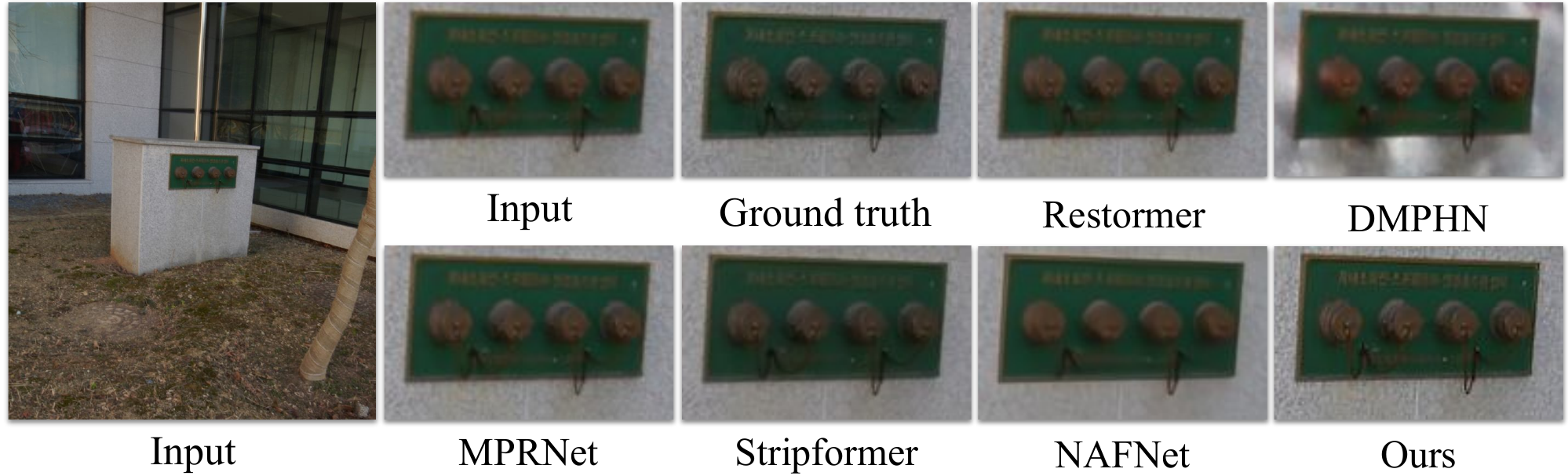}
\caption{Qualitative image deblurring comparison, with the image captured in daylight, between state-of-the-art models and our methods on RealBlur \citep{Realblur} dataset. Best viewed on a high-definition monitor when zoomed in}
\label{fig:RealBlur1}
\end{figure}
\begin{figure}[h]
\centering
\includegraphics[width=\linewidth]{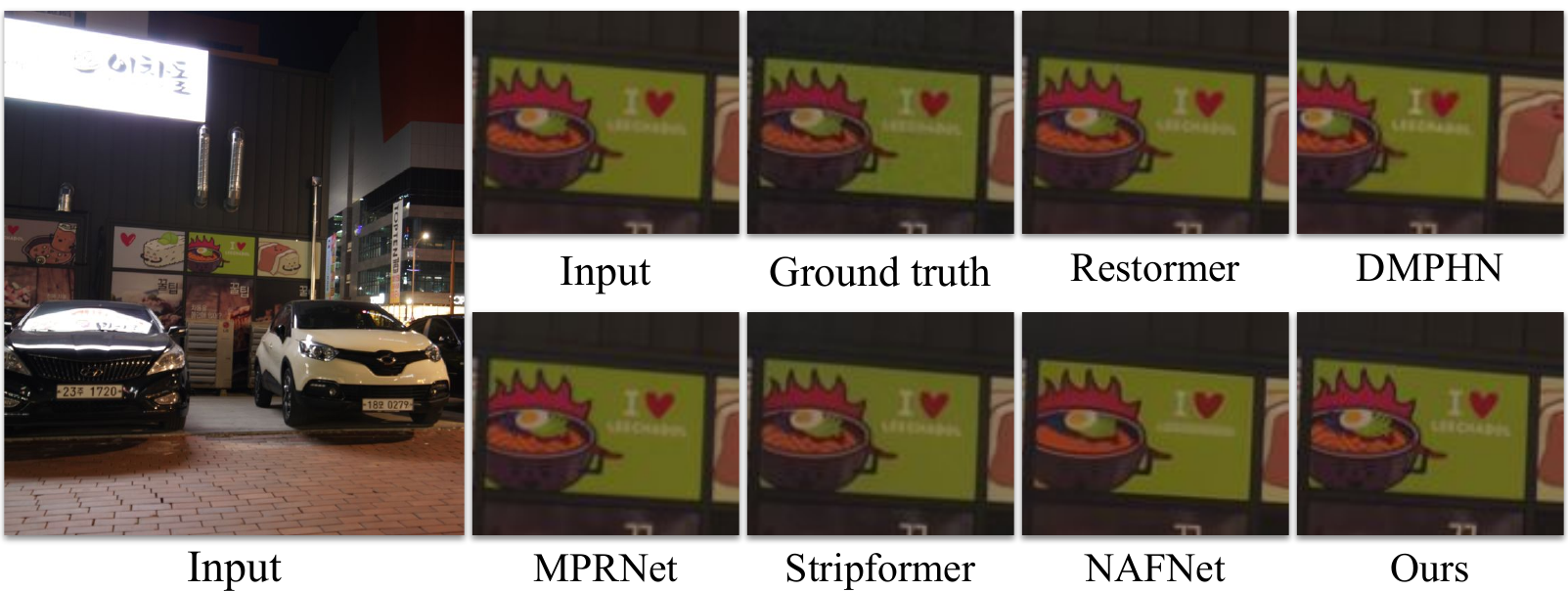}
\caption{Qualitative image deblurring comparison, with the image captured at night, between state-of-the-art models and our method on the RealBlur \citep{Realblur} dataset. Best viewed on a high-definition monitor when zoomed in}
\label{fig:RealBlur2}
\end{figure}
\begin{figure}[h]
\centering
\includegraphics[width=\linewidth]{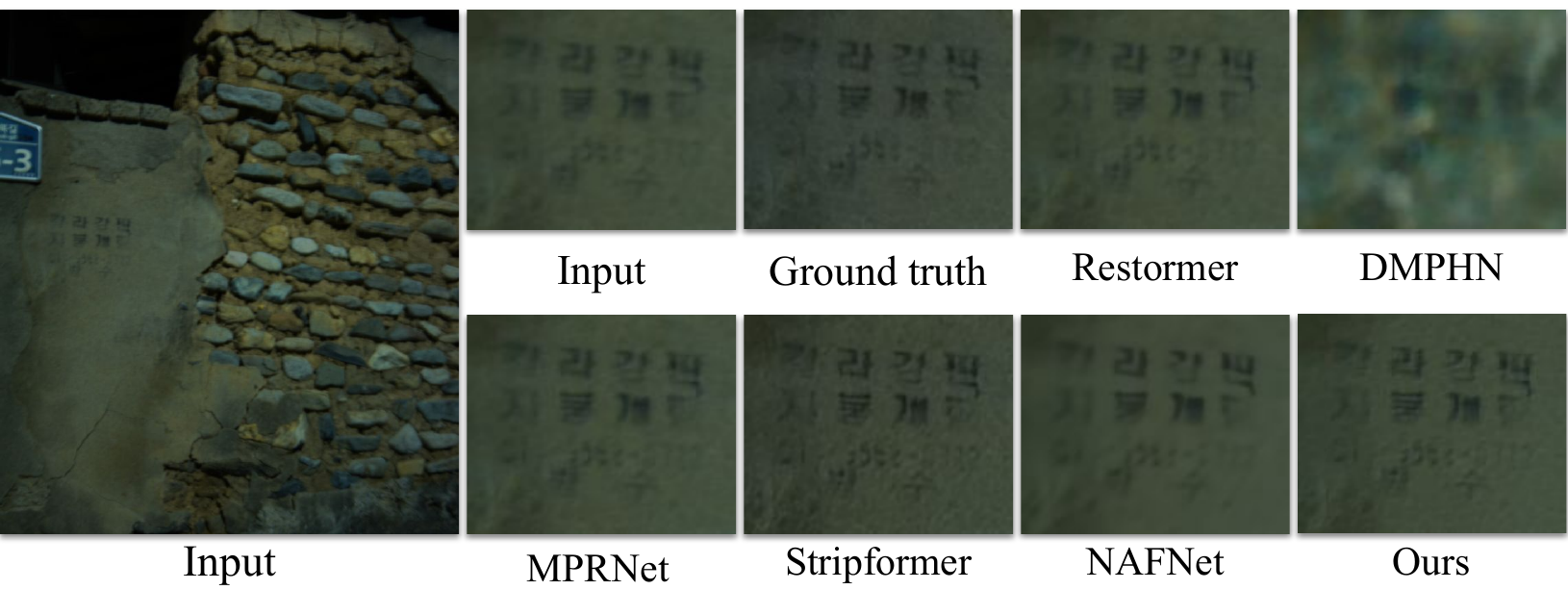}
\caption{Qualitative image deblurring comparison, with the image captured in the night, between state-of-the-art models and our methods on RealBlur \citep{Realblur} dataset. Best viewed on a high-definition monitor when zoomed in}
\label{fig:RealBlur3}
\end{figure}
\begin{figure}[h]
\centering
\includegraphics[width=\linewidth]{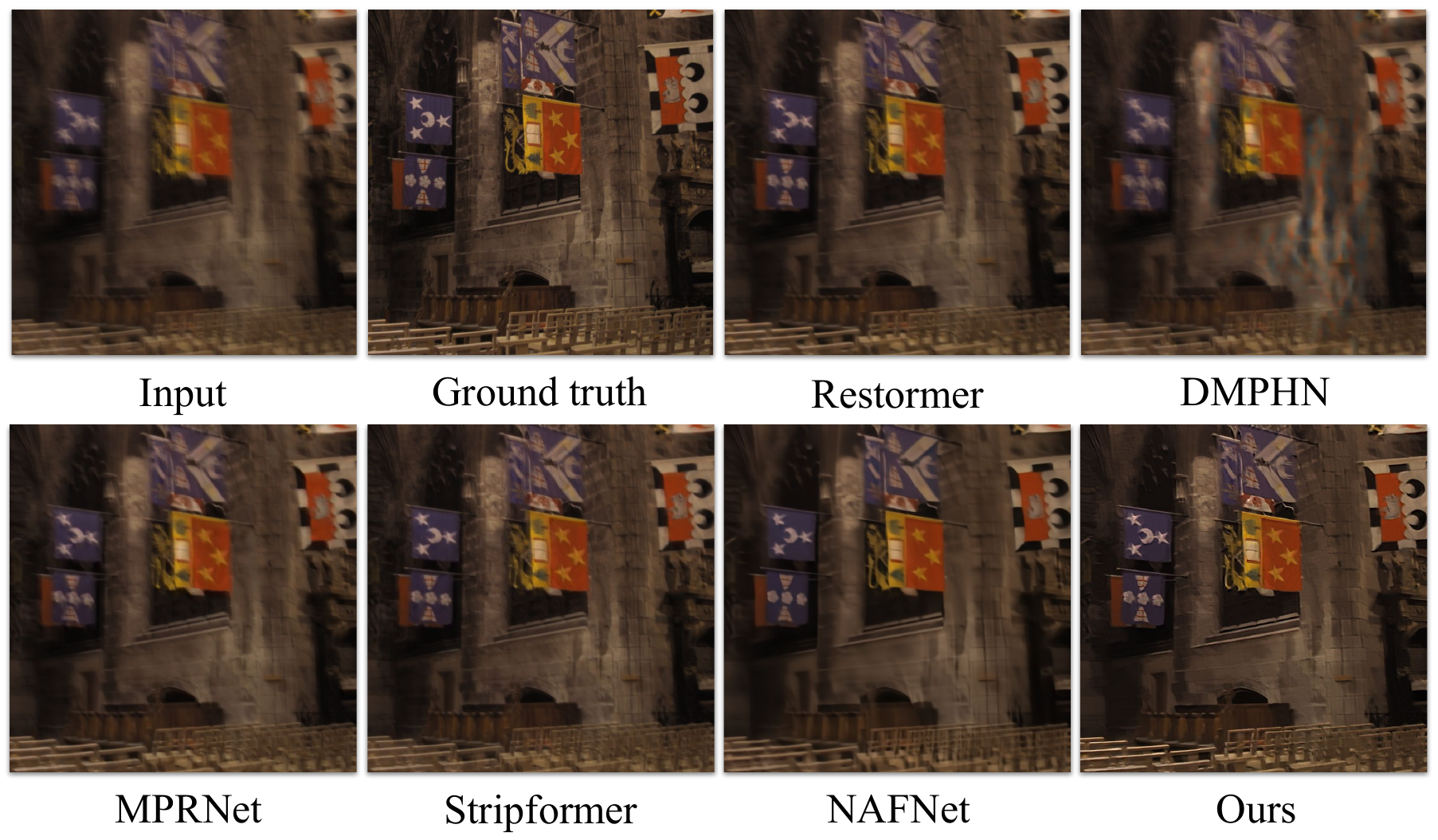}
\caption{Qualitative image deblurring comparison between state-of-the-art models and our methods on Kohler \citep{kohler}. Best viewed on a high-definition monitor when zoomed in}
\label{fig:kohler1}
\end{figure}
\begin{figure}[h]
\centering
\includegraphics[width=\linewidth]{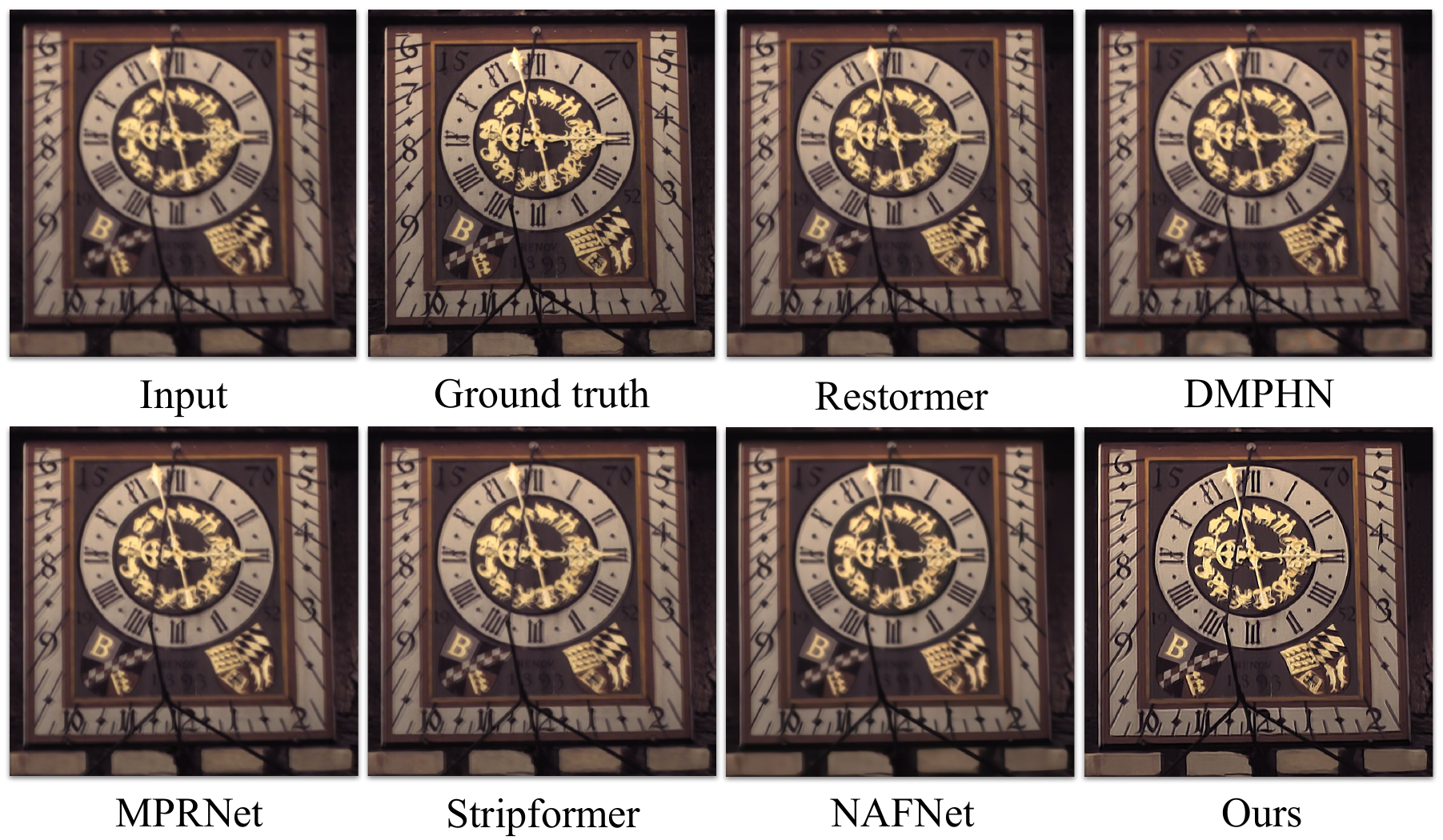}
\caption{Qualitative image deblurring comparison between state-of-the-art models and our methods on Kohler \citep{kohler}. Best viewed on a high-definition monitor when zoomed in}
\label{fig:kohler2}
\end{figure}

In Figure \ref{fig:Gopro} and \ref{fig:RealBlur1}, we evaluate our method in comparison to other state-of-the-art image deblurring techniques by utilizing daylight images from the RealBlur and GoPro datasets. The precision and clarity of images that have been affected by motion blur can be restored using our method. The images generated by our method are more visually convincing and sharper than those produced by other deblurring techniques, as illustrated in the figures. In Figure \ref{fig:RealBlur2} and \ref{fig:RealBlur3}, we evaluated the performance of our model against that of existing image deblurring methods by utilizing photographs captured at night from the RealBlur dataset. The strategy we employed consistently outperformed the other models in the retrieval of text that was more easily understandable and readable. Our method was found to be capable of producing images that were occasionally more detailed than the actual ground truth, in addition to being aesthetically appealing. Our methodology manages challenging deblurring tasks, as evidenced by the sharpness it achieves, particularly in low-light conditions. Most of the deep learning image deblurring models frequently employ benchmark datasets such as RealBlur, HIDE, and GoPro for training and testing. Figures \ref{fig:kohler1} and \ref{fig:kohler2} demonstrate that we expanded our evaluation to encompass the Kohler dataset following a thorough evaluation of our methodologies using these standard datasets. The complex blur patterns in this dataset serve as an excellent test scenario for evaluating the accuracy of picture deblurring algorithms. In contrast to competing models, ours demonstrated superior performance on the Kohler dataset.
\begin{table*}
\centering
\caption{Runtime, peak GPU memory, and throughput analysis}
\label{tab:runtime_memory_throughput}
\begin{tabular}{lccccc}
\toprule
\textbf{Method} & \textbf{Resolution} & \textbf{Time (s)} & \textbf{Peak GPU Mem (GB)} & \textbf{Images/min} \\
\midrule
Restormer (ViT only)       & 720p  & 1.13 & 23.00 & 54.5 \\
FFT-ReLU (standalone)      & 720p  & 0.90 & 0.75  & 66.7 \\
\textbf{Ours (ViT+FFT-ReLU)} & 720p  & \textbf{2.03} & \textbf{23.00} & \textbf{30.0} \\
\midrule
Restormer (ViT only)       & 1080p & 2.10 & 23.00 & 28.6 \\
FFT-ReLU (standalone)      & 1080p & 1.44 & 0.90  & 42.9 \\
\textbf{Ours (ViT+FFT-ReLU)} & 1080p & \textbf{3.54} & \textbf{23.00} & \textbf{17.1} \\
\bottomrule
\end{tabular}
\end{table*}

\subsubsection{Human Visual Preference Evaluation}
\label{HCI}
\begin{figure}[h]
    \centering
    \includegraphics[width=\linewidth]{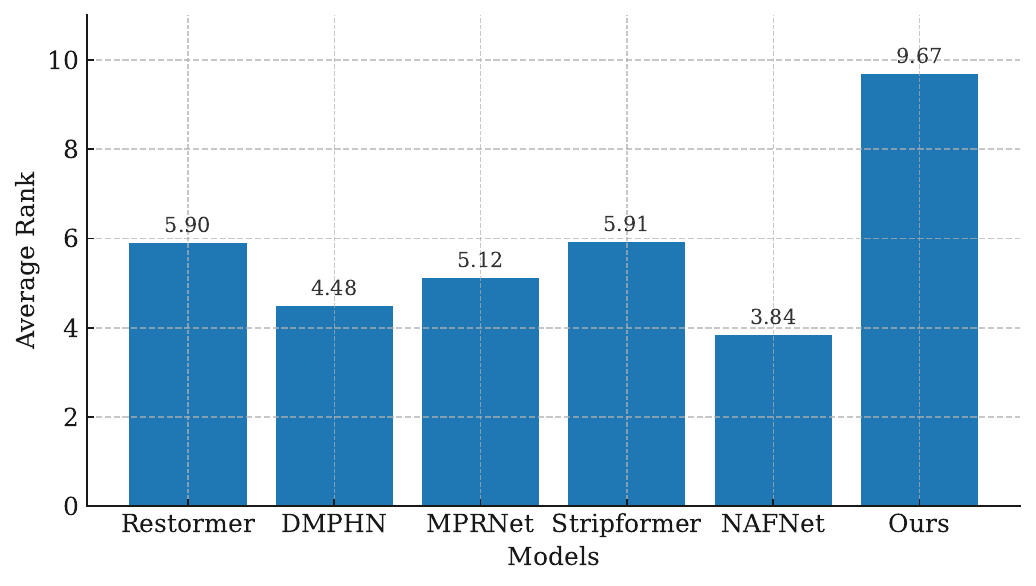}
    \caption{Histogram showing the ranking of each model in the Human Visual Preference Evaluation.}
    \label{test_result}
\end{figure}
In order to further validate our methodology, we conducted a Human Visual Preference Evaluation study to compare the output images of our model to those of extant SOTA image deblurring methods. This evaluation assesses the subjective visual quality of the photographs. While SSIM and PSNR are important quantitative metrics, they do not capture all of the nuances of human visual perception. Therefore, we concealed the method used to create each picture from the participants so they could compare them. The results of this anonymous interaction study allowed us to measure the impact our images had on human viewers' emotions. The results proved that our method consistently generated high-quality images that were visually appealing and often preferred over those generated by other algorithms.

\begin{enumerate}
    \item \textbf{Type of Study:} Using an online survey, we assessed the visual quality of deblurred photographs generated by various algorithms. DMPHN \citep{zhang2022eventDMPHN}, Stripformer \citep{Tsai2022Stripformer}, MPRNet \citep{Zamir2021MPRNet}, NAFNet \citep{chu2022nafssrNafnet}, and Restormer \citep{Restormer} were among the benchmark models employed to deblur a base picture in conjunction with our recommended model. In the survey, output images from each model were arbitrarily arranged in a 3x2 grid without labels. Ten distinct sets of these image grids were distributed to participants, who were instructed to evaluate each set on a scale of 1 to 10, with higher scores indicating superior quality.
    \item \textbf{Participants:} Volunteer participants were randomly selected without educational prerequisites. This approach ensured that the evaluators were drawn from a diverse range of backgrounds, thereby providing a more comprehensive understanding of the reception of the images. A total of 76 individuals participated in the survey.
\end{enumerate} 
Based on a scale of 1 to 10, Figure \ref{test_result} shows the typical rankings of the models that were under consideration for the survey. Our model had the highest average score (over 9.67 out of 10) among all the models examined. The assessment not only demonstrated that our method was effective in generating high-quality visual outputs, but it also demonstrated that it had practical applications in the real world, where the ultimate objective is to produce realistic and sharp visuals.

\subsubsection{Perceptual Improvement vs. Quantitative Gain}

While our model demonstrates performance gains in PSNR and SSIM, these improvements might seem small. However, we emphasize that image deblurring is often perceptually under-evaluated by these pixel-wise metrics. As seen in Figures 2–7, our method consistently restores sharper edges and preserves fine texture factors that are not always reflected in PSNR or SSIM. As per the Human Visual Preference Evaluation (Section \ref{HCI}), our method was preferred by a wide margin, receiving an average score of 9.67 out of 10 across 76 participants. This highlights the importance of perceptual clarity and sharpness in practical applications like photography. As such, we position our hybrid model not as a purely quantitative optimization, but as a practically superior restoration method, achieving consistent qualitative improvement that aligns better with human perception than traditional metrics.

\subsubsection{Runtime and Memory Analysis}

As shown in Table \ref{tab:runtime_memory_throughput}, Restormer alone requires 1.13s and 23 GB at 720p, increasing to 2.1s and 23 GB at 1080p. In contrast, the FFT-ReLU stage is extremely lightweight, completing in 0.9s - 1.44s with less than 1 GB of memory. When combined, our hybrid ViT+FFT pipeline runs in 2.03s (720p) and 3.54s (1080p) while keeping peak memory essentially unchanged at 23 GB, since the FFT stage adds negligible overhead. This analysis highlights that our frequency-domain module delivers substantial perceptual improvements (Section \ref{HCI}) at minimal additional runtime and memory cost. Crucially, FFT-ReLU scales with near-linear memory, in contrast to the super-linear growth of transformer layers. This makes our method more practical for high-resolution image restoration.

%% file: sec/5_conclusion.tex
\section{Conclusions and Future Work}
\label{Conclusions}
In this paper, we introduce a novel method of image deblurring that makes use of the computational efficacy of the Vision Transformer in conjunction with the Fast Fourier Transform (FFT) and ReLU sparsity approach. The integration of sparsity concepts with frequency domain processing yields enhanced deblurring performance and sharper images. Experiments show that our proposed method achieves impressive performance on various datasets. From the conducted human anonymous interaction survey, it can be said that our approach produces output that is more visually convincing than the output produced by other image deblurring methods. Our method is highly practical in real-world applications, as it consistently generates high-quality images that resonate with human perception and excels on common benchmarks. We position our work not as a single model, but as a dual-domain restoration paradigm, and future research can integrate lightweight ViTs or other backbones, as our paradigm is not tied to ViTs.